\documentclass[conference,compsoc]{IEEEtran}

\usepackage[T1]{fontenc}
\usepackage{graphicx}
\usepackage{tikz}
\usepackage{amsmath}
\usepackage{booktabs}
\usepackage{multirow}
\usepackage{color,soul}
\usepackage{algorithm,float}
\usepackage{amsfonts}
\usepackage[noend]{algpseudocode}
\usepackage{comment}
\usepackage{enumitem,kantlipsum}
\usepackage{textcase}
\usepackage{makecell}
\usepackage{url}
\usepackage{hyperref}
\usepackage{subfig}
\usepackage{seqsplit}
\usepackage{todonotes}
\usepackage{orcidlink}
\newtheorem{definition}{Definition}

\begin{document}
\makeatletter
\renewcommand{\paragraph}[1]{\par\medskip\noindent\textbf{#1}\par\nobreak\smallskip}
\makeatother

\title{DEFEAT: Stitching Fragmented File I/O Contexts for Early Ransomware Detection}

\author{
\IEEEauthorblockN{Muhammad Ejaz Ahmed}
\IEEEauthorblockA{CSIRO Technology\\
Sydney, Australia\\
ejaz.ahmed@csiro.au}
\and
\IEEEauthorblockN{Hyoungshick Kim}
\IEEEauthorblockA{Sungkyunkwan University\\
Suwon, South Korea\\
hyoung@skku.edu}
\and
\IEEEauthorblockN{Mohsen Ali Alawami}
\IEEEauthorblockA{Hankuk University of Foreign Studies\\
Seoul, South Korea\\
mohsencomm@hufs.ac.kr}
\and
\IEEEauthorblockN{Alsharif Abuadbba}
\IEEEauthorblockA{CSIRO Technology\\
Sydney, Australia\\
sharif.abuadbba@csiro.au}
\and
\IEEEauthorblockN{Seyit Camtepe}
\IEEEauthorblockA{CSIRO Technology\\
Sydney, Australia\\
seyit.camtepe@csiro.au}
\and
\IEEEauthorblockN{Surya Nepal}
\IEEEauthorblockA{CSIRO Technology\\
Sydney, Australia\\
surya.nepal@csiro.au}
\and
\IEEEauthorblockN{Junaid Qadir}
\IEEEauthorblockA{Qatar University\\
Doha, Qatar\\
jqadir@qu.edu.qa}
}

\maketitle

\begin{abstract}
Ransomware increasingly fragments its file operations across temporary and intermediate files, scattering the semantic context that links individual I/O events to an overarching encryption campaign. This fragmentation defeats existing detectors that reason over isolated file streams, whether pattern-based methods that match rigid event sequences or learning-based methods that require accumulating statistical evidence across many files.
We present \textsc{Defeat}, a framework that reconstructs this fragmented, scattered context by grouping causally related file events into \emph{File Event Gadgets} (FEGs), semantically coherent units that capture the full intent behind sequences of file operations spanning multiple dynamically created files. Unlike provenance graphs (system-wide causal graphs that record relationships among all OS entities, such as processes, files, sockets, and registry keys, across the entire system), FEGs are scoped to the file-operation context of a single user asset, enabling lightweight, targeted analysis without whole-system instrumentation. Each FEG is modelled as an attributed control flow graph (ACFG) and embedded via a graph neural network for unsupervised clustering, enabling analysts to label entire behavioural clusters rather than individual samples, reducing annotation effort by 94\%.
Evaluated on a corpus of 97{,}816{,}471 file I/O events spanning 67 ransomware families, \textsc{Defeat} achieves 99.2\% detection accuracy and outperforms state-of-the-art methods UNVEIL, RWGuard, and Peeler by 6.57--7.56 percentage points. The framework operates at the granularity of a single file encryption: because each ACFG represents exactly one FEG (one user asset context), a cluster label can be assigned as soon as the first file's encryption sequence completes, enabling detection at the first encrypted file. Beyond detection, \textsc{Defeat} uncovered 165 previously unreported malicious file I/O patterns, including write-before-read, multi-rename chain, and delete-then-recreate variants, from which production-grade threat-detection rules were derived and deployed. The dataset is available at \url{https://github.com/ejaz629/defeat}.
\end{abstract}


\section{Introduction} \label{sec:introduction}

Ransomware remains among the most damaging classes of cyber threats. The Australian Signals Directorate reported a 3\% increase in ransomware-related incidents in 2024-25~\cite{asd2025actr}, while global ransom payments exceeded USD~1.1 billion in 2023 alone~\cite{chainalysis2024}. Modern ransomware operators no longer rely on monolithic encryption routines; instead, they deliberately fragment file operations across auxiliary files, temporary buffers, and system-mediated I/O paths to defeat monitoring tools that
reason over individual file streams in isolation~\cite{kharaz2016unveil,ahmed2021peeler}.

\begin{figure}[t]
  \centering
  \includegraphics[width=0.98\linewidth,clip]%
    {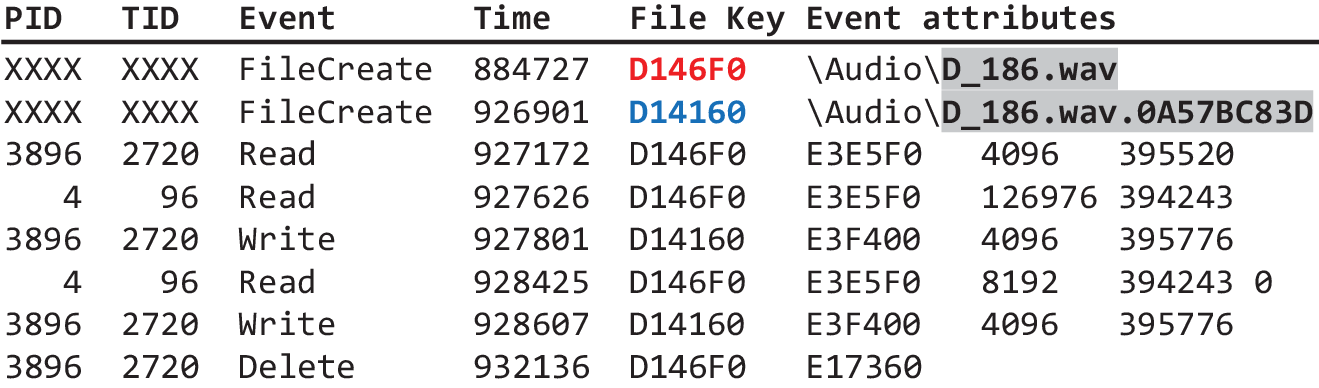}
  \caption{File I/O events generated by InfinityCrypt
    ransomware. Red and blue indicate different file keys
    used for the original and encrypted files,
    respectively.}
  \label{fig:strat2}
\end{figure}

Figure~\ref{fig:strat2} illustrates this evasion in practice. The InfinityCrypt variant assigns its original user file (\texttt{D\_186.wav}) and encrypted output (\texttt{D\_186.wav.0A57BC83D}) to separate file keys (\texttt{D146F0} and \texttt{D14160}). Viewed independently, each key reveals only benign-looking operations: creation, reads, writes, or deletion. The malicious intent emerges only when both keys are correlated: reading the original, writing encrypted content into a new file, and deleting the source. A system process (\textit{PID~4}) performing the read further obscures the attack by making it appear as routine system activity. Our empirical study of 67 ransomware families confirms that this is not an isolated case: approximately 44\% create multiple auxiliary files, each exhibiting a distinct set of I/O events, rendering per-file analysis insufficient.

Existing defences fall into two broad categories, neither of which adequately addresses this fragmentation. \emph{Pattern-based} methods such as UNVEIL~\cite{kharaz2016unveil} and Redemption~\cite{kharraz2017redemption} match hand-crafted file I/O sequences and can flag an attack as early as the first encrypted file; however, they treat each file in isolation and rely on manually curated signatures that erode as adversaries mutate their access patterns~\cite{continella2016shieldfs,scaife2016cryptolock,huang2017flashguard,milajerdi2019poirot}. \emph{Learning-based} methods such as Peeler~\cite{ahmed2021peeler} and RWGuard~\cite{mehnaz2018rwguard} generalize better by modelling statistical distributions of event features, yet they require accumulating enough observations before reaching a verdict, allowing multiple files to be encrypted before detection~\cite{sgandurra2016automated,zhao2019tee,alahmadi202299,al20170}. In short, pattern-based approaches offer speed but poor adaptability; learning-based approaches offer adaptability but cannot guarantee early-stage detection. Crucially, both treat
file events in isolation, leaving the contextual fragmentation exploited by modern ransomware unaddressed.

To bridge this gap, we introduce DEFEAT (\textbf{\underline{DE}}tecting ransomware through contextual \textbf{\underline{F}}ile \textbf{\underline{E}}vent \textbf{\underline{A}}nalysis and unsupervised clus\textbf{\underline{T}}ering), a framework that
reconstructs the fragmented context of file operations before classification. DEFEAT introduces \emph{file event gadgets} (FEGs), semantically coherent units that capture related I/O events across multiple dynamically created files sharing a common operational context. FEGs are modelled as attributed control flow graphs (ACFGs) encoding the behavioural structure of each context. These graphs are embedded via unsupervised graph-level representation learning into a vector space that preserves intra- and inter-graph proximity,
enabling clustering of similar behavioural patterns. Analysts then label a small number of representative clusters rather than individual files, and the resulting cluster-level labels propagate to all constituent samples. The behavioural patterns derived from malicious clusters yield a richer set of attack signatures than prior work, which relied on only a handful of manually identified patterns~\cite{kharaz2016unveil,kharraz2015cutting,ahmed2021peeler}. Fragmentation motivates context reconstruction but is not a precondition for detection: a non-fragmenting sample yields a single-file FEG, classified through the identical pipeline. Such single-file encryption accounts for 56\% of our corpus and is included in all reported results.
Our contributions are as follows.

\begin{itemize}[leftmargin=*, label={--}]

\item \textit{File Event Gadgets (FEGs).}  We propose file event gadgets (FEGs), a novel abstraction that captures semantically related file I/O events across multiple files. From 97{,}816{,}471 file I/O events spanning 67 ransomware families and benign applications, we extract 675{,}140 FEGs, a compact yet comprehensive representation that enables holistic analysis of fragmented ransomware activity.

\item \textit{Graph-Based Behavioural Modelling.} FEGs are transformed into attributed control flow graphs and embedded via unsupervised graph representation learning. Clustering these embeddings into 256 groups reveals 165 previously unreported malicious behavioural patterns, demonstrating generalization across diverse ransomware families.

\item \textit{Efficient Detection with Reduced Analyst Effort.} Cluster-level labelling reduces human annotation effort by 94\% compared to per-file analysis, while maintaining over 99\% detection accuracy-- outperforming UNVEIL, RWGuard, and Peeler on the same dataset.

\item \textit{Generalization to Unseen Threats.} Without retraining, DEFEAT correctly identifies malicious functionality in 49~samples from 24 previously unseen ransomware families (Table~\ref{tab: dataset stats}), demonstrating strong generalization beyond training families.

\item \textit{Large-Scale Dataset and Tool Release.} We develop a Windows kernel-level module leveraging Event Tracing for Windows (ETW) to capture fine-grained file events unavailable via conventional tools such as Sysmon. The resulting dataset, 97{,}816{,}471 events from 67 families, is publicly available at \url{https://github.com/ejaz629/defeat}.

\end{itemize}


\section{Background and Motivation} \label{sec:background}
In this section, we give an overview of the Event Tracing for Windows (ETW) module in Windows OS and provide our key observations to detect ransomware attacks in terms of file I/O event patterns.
\subsection{File I/O Schema in ETW} \label{ssec:schema detail}
Modern Windows operating systems offer Event Tracing for Windows (ETW)~\cite{ETW} as a low-overhead, kernel-level instrumentation framework that records fine-grained system events. Among these, file I/O events are particularly valuable for security monitoring, as they capture the full lifecycle of file interactions, from creation and opening to subsequent reads, writes, renames, and deletions. As shown in Table \ref{tab:systemevents}, these events expose both common attributes (e.g., process identifier, thread identifier, and precise timestamp) and operation-specific attributes (e.g., file object handles, file names, access flags, and I/O size). Such detailed metadata allows us to accurately link file operations to their originating processes and reconstruct file access patterns over time. This capability is critical for building reliable behavioral contexts, enabling the detection of stealthy file modifications, unauthorized access attempts, or coordinated malicious activities that may otherwise evade traditional monitoring approaches.

\begin{table}[!th]
    \centering
    \caption{File events schema in ETW.}
    \resizebox{0.95\linewidth}{!}{
    \begin{tabular}{lll} 
    \toprule
         \multirow{2}{*}{\textbf{File I/O event}} & \multicolumn{2}{c}{\textbf{Event schema}} \\ \cmidrule{2-3}
           & \textbf{Common attributes} & \textbf{File I/O event-specific attributes}  \\
          \midrule
         Read, Write & PID, TID, Timestamp & \multicolumn{1}{l}{FileKey, FileObject, IoSize, IoFlags}\\ \cmidrule{1-3}
         Rename, Delete & PID, TID, Timestamp & \multicolumn{1}{l}{FileKey, FileObject}\\ \cmidrule{1-3}
         FileCreate, FileDelete & PID, TID, Timestamp & \multicolumn{1}{l}{FileObject, FileName} \\ \cmidrule{1-3}
         Create &  PID, TID, Timestamp & \multicolumn{1}{l}{FileObject, OpenPath} \\
          \bottomrule
    \end{tabular}
    }
\label{tab:systemevents}
\end{table}



\subsection{Ransomware File Encryption Patterns} \label{ssec:ransowmare strategies}
Typically, ransomware encrypts a user file through the following four steps: 1) access the file (\textit{access}); 2) read the content of the file (\textit{read}); 3) write the encrypted content to a temporary memory or new file (\textit{write}); and 4) overwrite/delete (depending on the strategy) the user's original file (\textit{overwrite/delete}). 

\begin{figure}[!th]
    \centering
    \includegraphics[width=0.98\linewidth,clip]{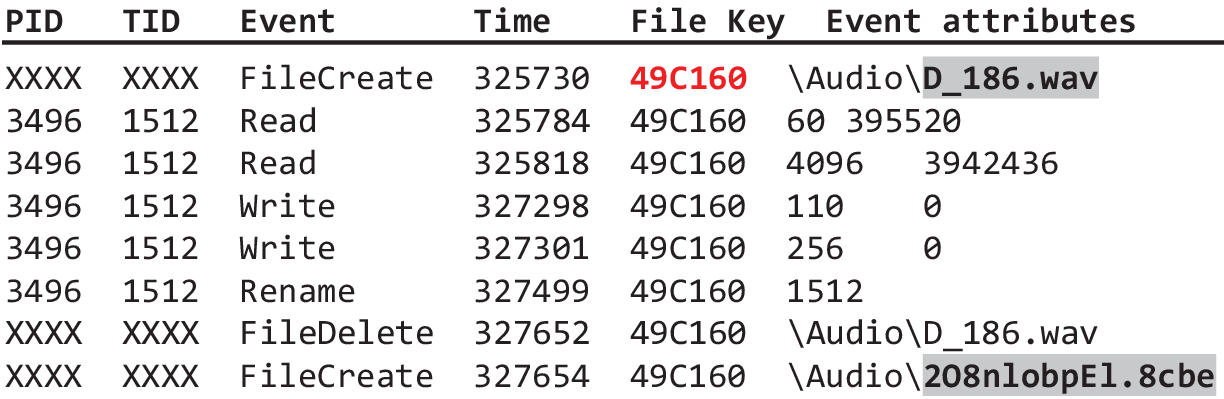}
    \caption{File I/O events generated by Cerber ransomware. The red color represents a single, one-of-a-kind file key, while the grey color indicates the files that are included in the encryption process.}\label{fig:strat1A}
\end{figure}

We examine a prevalent file I/O access pattern used by ransomware to encrypt user files. As illustrated in Figure~\ref{fig:strat1A}, this strategy unfolds in four steps: (1) access, the ransomware sample accesses a file (\textit{D\_186.wav}) with the \textit{FileCreate} event; 2) in the \textit{read} step, the ransomware sample reads the content of \textit{D\_186.wav} with the two \textit{Read} events; 3) in the \textit{write} step, the ransomware sample writes the encrypted content to the same file with the two \textit{Write} events; and 4) in the \textit{overwrite} step, the file is finally renamed with the \textit{Rename}, \textit{FileDelete}. The content of the original file \textit{D\_186.wav} assigns a new name \textit{2O8nlobpEl.8cbe}. 
However, the number of \textit{FileKey}s may vary during the encryption process, and hence it becomes challenging to identify and combine contextually-related file I/O events from multiple files. Our empirical results show that around 44\% of ransomware families rely on more than one file (\texttt{file key}) for encryption. For that reason, we have introduced the concept of the event gadget to collect all events, discussed in Section~\ref{ssec:event gadget extraction}.
Our analysis of 67 ransomware families shows that $\approx$56\% (e.g., Cerber, Keypass, TeslaCrypt, VirLock, GandCrab, GlobeImposter) employ this strategy with a single persistent \textit{FileKey}. However, ~44\% use multiple keys during encryption, complicating the task of linking related file I/O events.

\subsection{Threat Model}
We consider user-space ransomware that executes with the privileges of the victim user (including administrative, but not kernel-level, privileges) and attempts to encrypt user files. We assume the OS kernel and the ETW infrastructure are trusted, and that the adversary cannot disable, tamper with, or forge kernel-level ETW events. This is the standard assumption of prior host-based detectors, including UNVEIL~\cite{kharaz2016unveil}, RWGuard~\cite{mehnaz2018rwguard}, and Peeler~\cite{ahmed2021peeler}. An adversary with kernel privileges can disable any in-OS defence, including all baselines, and is out of scope; complementary protections (e.g., hypervisor- or TEE-based monitoring~\cite{zhao2019tee}) address that setting. Within scope, we assume a strong adversary with full control over the order and nature of the file I/O events its process generates, subject only to performing encryption (read original, write ciphertext, delete or overwrite source); we evaluate this adversary in Section~\ref{ssec:adversarial}.

\section{System Design} \label{sec:system design}
\begin{figure*}[!th]
  \centering
  \includegraphics[width=0.9\textwidth,clip]{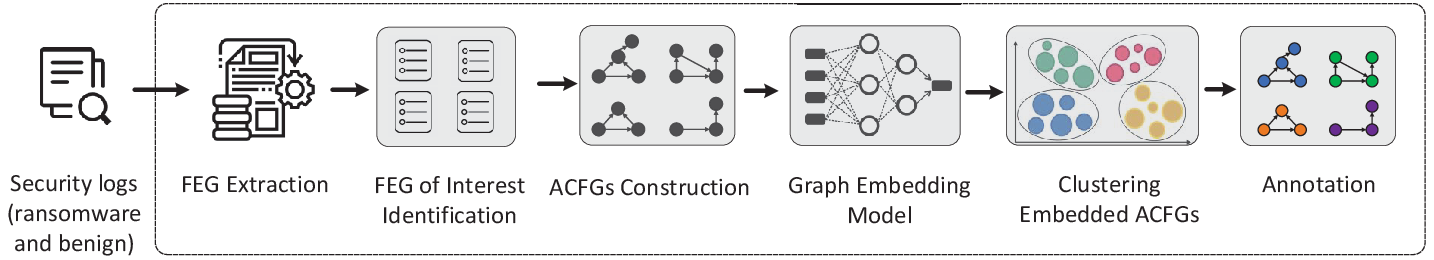}
  \caption{Overview of DEFEAT. \textit{The file I/O events from security logs are used to extract FEGs, which are then filtered to get the FEGs of interest and converted to attributed graphs (ACFGs); clusters of graph embeddings computed from the ACFGs are then provided to an analyst for collective labeling}.}
\label{fig:system design}
\end{figure*}

In this section, we present a comprehensive overview of the DEFEAT pipeline, illustrated schematically in Figure~\ref{fig:system design}. Our approach introduces the concept of a \textit{file event gadget} (FEG) to represent the behavior of an executing program from low-level file I/O events within their execution context. An FEG is a collection of several low-level file I/O events performed on one or more files, designed to connect fragmented contexts and provide meaningful information about the overall intention of file operations (Section~\ref{ssec:event gadget extraction}).

To focus on potentially malicious activities, we filter the obtained FEGs to extract FEGs of interest (Section~\ref{ssec:foi identification}). DEFEAT considers file events typically employed by ransomware to encrypt user files, such as file access, read, write, delete, or rename operations. These filtered FEGs are then transformed into directed-labeled graphs, called Attributed Control Flow Graphs (ACFGs), to capture the control flow structure of file operations specific to ransomware behaviors (Section~\ref{ssec:feg to graphs}).

For unsupervised and inductive learning of graph representations, we employ UGRAPHEMB~\cite{bai2019unsupervised} to construct a graph embedding model from the ACFGs (Section~\ref{ssec:graph embedding model}). This model generates embeddings for each ACFG, which are subsequently used for clustering. The resulting clusters can be collectively annotated by human analysts, significantly reducing their workload and enhancing the efficiency of ransomware detection (Section~\ref{ssec:clustering model}).

\subsection{FEG Extraction} \label{ssec:event gadget extraction}
\begin{figure*}[h]
  \centering
  \includegraphics[width=0.9\textwidth,clip]{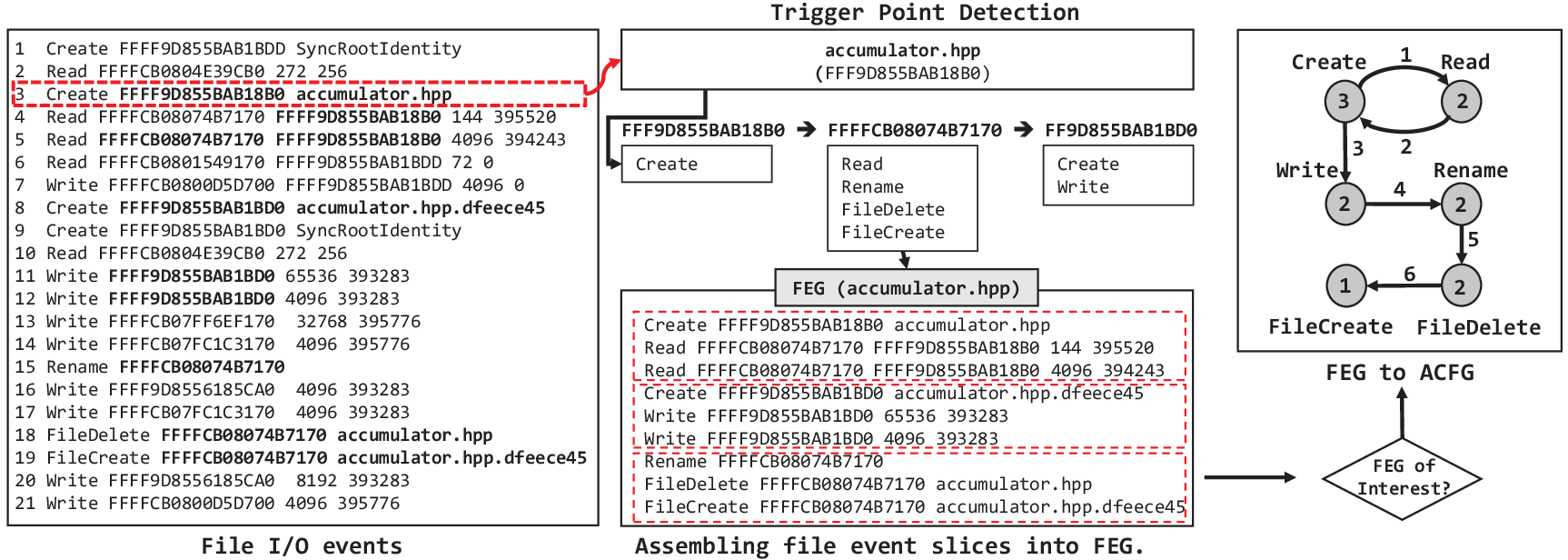}
  \caption{Overview of FEG extraction: the incoming stream of file I/O events, (left). To combine contextually-related file I/O events from multiple files (i.e., \texttt{.hpp} and \texttt{.dfeece45}), the file I/O events are stitched together with the keys of the three files, each with corresponding file I/O events, to construct an FEG, (middle). Finally, the FEG of interest is converted to graphs (right).}
\label{fig:file event gadget extraction}
\end{figure*}

Figure~\ref{fig:file event gadget extraction} illustrates the process of extracting FEGs from low-level system events. When an application accesses a file, specific file I/O events such as \textit{FileCreate} or \textit{Create} are generated. These events are called the \textit{trigger point}. DEFEAT is triggered by such file access events, and it begins profiling file I/O events associated with the user/system file while constructing an FEG for that particular file.
As shown in Figure~\ref{fig:file event gadget extraction}, the \textit{FileKey} from the \textit{trigger point} event is extracted and used to correlate other events sharing the same context. Since only the \textit{Create}, \textit{FileCreate}, and \textit{FileDelete} events have the \textit{FileName} attribute (Table~\ref{tab:systemevents}), DEFEAT uses the \textit{FileKey} and \textit{FileObject} attributes of other events (e.g., \textit{Read}, \textit{Write}, \textit{Rename}, and \textit{Delete}) to combine and construct a FEG. The FEG construction process for the file ``\texttt{accumulator.hpp}'' is demonstrated in Figure~\ref{fig:file event gadget extraction}, where three distinct \textit{FileKeys} combine all relevant file I/O events into the FEG. 


\begin{definition}[File Event Gadget]
  \label{def:feg}
  Let $\mathcal{E} = \langle e_1, e_2, \ldots \rangle$ be the ordered stream of
  file I/O events observed by the ETW monitor. A \emph{File Event Gadget} (FEG)
  is a tuple $\mathcal{G} = \langle \mathcal{F},\, E_\mathcal{G} \rangle$, where

  $\mathcal{F} = \{f_1, f_2, \ldots, f_k\}$ is the set of related files sharing
  a common operational context, and $E_\mathcal{G} \subseteq \mathcal{E}$ is the

  ordered set of file I/O events associated with files in $\mathcal{F}$.

  Two files $f_i, f_j \in \mathcal{F}$ are considered \emph{contextually
  related}
  if they satisfy at least one of the following conditions:
  \begin{enumerate}[label=(\roman*)]
    \item \textbf{Shared FileKey:} they share the same kernel-assigned file
  handle
          identifier at some point during the observation window;
    \item \textbf{Shared FileObject:} they share the same kernel object pointer;
    \item \textbf{Filename-prefix match:} the filename of $f_j$ without its
          extension equals the filename of $f_i$ (e.g., \texttt{x.txt} and
          \texttt{x.txt.locked} belong to the same FEG);
    \item \textbf{Same-path temporal adjacency:} they reside in the same
  directory
          and their file I/O events overlap within the same process execution
  context.
  \end{enumerate}

  An FEG is \emph{initiated} by a trigger event (\texttt{FileCreate} or
  \texttt{Create}) on a user file and \emph{closed} when no new contextually
  related file events are observed within the active process context. The FEG
  boundary is therefore process-scoped and file-centric, distinguishing it from
  a provenance graph~\cite{milajerdi2019poirot}, which is system-wide and entity-agnostic.
  \end{definition}

\begin{algorithm}[h]
\scriptsize
\caption{File event gadgets (FEGs) extraction.}\label{algo:algo1}
\begin{algorithmic}[1]
\Statex \textbf{Input: } Incoming file I/O events.
\Statex \textbf{Output: } \textit{File event gadgets (FEGs)}
\Statex
\Statex \textbf{Stage 1: Detect \textit{trigger key} from file I/O events.}
\If{file I/O event is \textit{FileCreate} or \textit{Create}}
\State Create file event gadget for the user file in \textit{trigger key} file event.
\Statex
\Statex \textbf{Stage 2: Create file event gadgets.}
    \If{the \textit{FileName} of the \textit{event} is newly observed}
        \If{the \textit{FileName} without extension is already observed} \Comment{e.g., the file `myfile.txt.locked' belongs to the file event gadget of the file `myfile.txt,' and hence no need to create a new file event gadget for that file.}
        \State Add \textit{event} to \textit{FileName}'s file event gadget.
        \Else {
            \If{there exists a file (\textit{FileName}) against the \textit{FileKey} of the \textit{event}}
            \State Add \textit{event} to \textit{FileName}'s file event gadget.
            \Else{ Create a new file event gadget with the file name \textit{FileName}}
            \EndIf
        }
        \EndIf
    \EndIf
    \If{there exists a file event gadget for the file \textit{FileName}}
        \If{there exists another file against \textit{FileKey} and their \textit{paths} are also same}
        \State Add \textit{event} to the other \textit{FileName}'s file event gadget.
        \Else{\State Add \textit{event} to the original file's \textit{FileName}'s file event gadget.}
        \EndIf
    \EndIf
\Statex
\Statex \textbf{Stage 3: Add events to respective file event gadgets.}
\If{\textit{event} type is \textit{Read}, \textit{Write}, \textit{Rename}, or \textit{Delete}}
    \If{there exists a file \textit{FileName} against the \textit{FileKey}}
        \State Add \textit{event} to \textit{FileName}'s file event gadget.
    \EndIf
\EndIf
\EndIf
\end{algorithmic}
\end{algorithm}
Algorithm~\ref{algo:algo1} describes all steps for FEGs extraction. There are three main stages in FEGs extraction: 1) detect \textit{trigger point} file events, 2) create FEGs, and 3) add events to corresponding FEGs. As illustrated in the Algorithm~\ref{algo:algo1} (Step 1 -- 2), if \textit{Create} or \textit{FileCreate} events for a user file are observed, its \textit{FileName} and \textit{FileKey} attributes are extracted. The user file name is used to create a FEG to place corresponding file I/O events in the event gadget. The \textit{FileKey} is used to match against other file events and assemble those events in the created FEG. DEFEAT then checks if there already exists an FEG for the user file. If the FEG does not exist, DEFEAT also checks if the user file exists in the FEG database when the file's extension is removed. If the file exists, DEFEAT adds this event to the corresponding FEG (Step 3 -- 6). DEFEAT adds file events to their corresponding FEGs leveraging the \textit{FileName} or \textit{FileKey} attributes (Step 7 -- 9). These steps are necessary because a given user file may have different \textit{FileKeys} over time and vice versa. For example, Figure~\ref{fig:file event gadget extraction} demonstrates that there are three unique \textit{FileKeys} representing the FEG for the file \texttt{accumulator.hpp}. Moreover, to ensure that file I/O events are correctly placed in their corresponding FEGs, DEFEAT checks the \textit{path} of the user file (Step 10 -- 14). Finally, all other events (i.e., \textit{Read}, \textit{Write}, \textit{Rename}, and \textit{Delete}) are placed into their corresponding FEGs (Step 15 -- 17) to provide the context information about the operations in a FEG.

\subsection{FEGs of Interest Identification} \label{ssec:foi identification}
The rationale for using the ``FEGs of interest'' is to narrow down and focus on the behaviors that are typically observed when ransomware samples execute, such as file rewriting, renaming, and deletion, as demonstrated in previous studies~\cite{kharaz2016unveil,ahmed2021peeler,kharraz2017redemption}. To effectively identify FEGs of interest, DEFEAT relies on detecting a sequence of suspicious file I/O events (e.g., file access, read, write encrypted content, and file renaming or delete operations on a user's file) in each FEG returned from Algorithm~\ref{algo:algo1}.
Although the FEG may include the file I/O events mentioned above, it does not necessarily indicate a ransomware attack pattern. We have noticed that some harmless applications can generate file I/O patterns that resemble ransomware behavior. For instance, benign applications typically overwrite the Windows OS's Activation Tokens file (\textit{tokens.dat}), showing similar file I/O operations to those of ransomware. As a result, this could lead to false positives.
We have observed that occurrences of \texttt{tokens.dat} file overwrite are exceedingly infrequent, constituting less than 1\% of cases, as per our empirical analysis of benign applications presented in Table~\ref{table:Benign_applications}.

To extract the FEGs of interest, we consider FEGs that contain at least one \emph{write} event combined with at least one of \{read, delete, rename\}. This relaxes the strict four-event requirement to accommodate overwrite-only ransomware variants (e.g., those that encrypt in-place without a delete step), while still filtering purely read-only or creation-only FEGs that cannot represent an encryption workflow. Requiring all four event types (\textit{read}, \textit{write}, \textit{delete}, \textit{rename}) was evaluated empirically on our dataset and produced 2.3\% fewer FEGs of interest -- missing exactly the overwrite-in-place variants. The looser criterion captures 99.97\% of all ransomware FEGs while reducing benign FEG retention by only 0.04\%, confirming that the filter remains highly selective. These file I/O operations are the core actions required to encrypt a user file, consistent with prior studies~\cite{ahmed2021peeler,kharaz2016unveil,kharraz2017redemption,mehnaz2018rwguard}. Our approach is agnostic to the PID, which means that if another process generates file I/O events that are contextually relevant, they will be included in the same FEG. As illustrated in Figure~\ref{fig:strat2}, file I/O events generated by system processes (such as \texttt{explorer.exe} with PID=4) are merged into the same FEG.

\subsection{ACFGs Construction} \label{ssec:feg to graphs}
We transform the FEGs of interest into ACFGs to provide their control flow structures. 
The rationale is that adaptive attackers may reorder or inject file I/O events to break sequence-dependent detectors such as UNVEIL~\cite{kharaz2016unveil,kharraz2017redemption}. ACFGs address this by encoding the \emph{event-type transition graph} rather than the raw sequence: nodes represent event types (e.g., \texttt{Read}, \texttt{Write}, \texttt{Delete}) and directed edges represent observed transitions between types. Two structural properties remain stable under surface-level manipulation: (1)~the set of event types present (node set), which is determined by the encryption algorithm and cannot be removed without breaking functionality; and (2)~the dominant transition topology (e.g., Read$\to$Write always present in any encryption that reads plaintext and writes ciphertext). Injecting dummy events adds isolated or weakly connected nodes that the GNN embedding model weights less than the densely connected core, and reordering events alters edge direction but not the overall reachability structure. This design makes ACFGs structurally robust to the perturbation classes that defeat pattern-matching approaches. Note that ACFGs are built from runtime file I/O events, not from program code. Binary-level obfuscation (e.g., packing or control-flow flattening) changes the program's code structure but not the file operations it must perform to encrypt files, and therefore does not affect the ACFG.

DEFEAT constructs an ACFG from an FEG of interest by creating a node for the first event type (such as a Create or FileCreate event). If a new event type, such as Read, Write, or Delete, is observed, a node representing that event type (denoted as $n_t$) is created. A directed edge ($n_i$, $n_j$) is added when event $j$ is generated immediately after event $i$. We do not add an edge if two subsequent events are the same type of event. As shown in Figure~\ref{fig:file event gadget extraction}, the created node has two attributes: 1) file event type (above/below the circle in Figure~\ref{fig:file event gadget extraction}), and 2) the number of neighbor nodes (inside the circle in Figure~\ref{fig:file event gadget extraction}). Each ACFG provides the temporal characteristics of file I/O events on a file.

\subsection{Graph Embedding Model} \label{ssec:graph embedding model}
To convert ACFGs into embeddings, we apply the graph-level representation learning method known as UGRAPHEMB~\cite{bai2019unsupervised}. This technique embeds graphs into a vector space while maintaining the proximity relationships between them, ensuring that similar graphs have closer embeddings. UGRAPHEMB generates embedded graphs so that similar graphs are embedded closer to each other in feature space because their graph proximity distances are maintained for graph embedding. UGRAPHEMB originally used Graph Edit Distance (GED) as a graph proximity metric. However, because the GED optimization problem is known to be NP-Hard~\cite{bai2019unsupervised}, we replace it with a novel graph proximity metric based on the \emph{Laplacian spectral distance}: the $\ell_2$ distance between the sorted eigenvalue spectra of the normalized Laplacian matrices of two graphs. This metric is computable in $O(n^3)$ time for $n$-node graphs and captures global structural similarity -- graphs with similar connectivity patterns will have similar eigenvalue distributions -- making it well suited to comparing small ACFGs (typically 3--8 nodes). In UGRAPHEMB, the trained model is considered a function that receives any graph as input and transforms it into an embedding using a graph-level embedding generation mechanism called Multi-Scale Node Attention (MSNA). Node embeddings within MSNA are computed using Graph Isomorphism Networks (GIN)~\cite{xu2018powerful}, a maximally expressive message-passing architecture that aggregates neighbor features via learnable MLP layers, ensuring permutation-invariant, inductive graph representations. 

Given ACFGs, UGRAPHEMB first generates a set of node embeddings while maintaining their inductivity and permutation invariance. For node embedding, it relies on the state-of-the-art neighbor aggregation method Graph Isomorphism Network (GIN)~\cite{xu2018powerful}.
The intuition is to embed the data points in a low dimensional space such that their pairwise proximity distances are preserved, e.g., via minimizing the loss function:
\begin{align}
\mathcal{L}(h_i, h_j,d_{ij}) = \left(\lVert h_i - h_j\rVert_2^2 - d_{ij}\right)^2,
\label{Eq3}
\end{align}
where $h_i$ and $h_j$ represent the embeddings of datapoints $i$ and $j$, respectively, and $d_{ij}$ represents their distance.
After training, the learned neural network model can be applied to any graph, and the graph-level embeddings can be used on several downstream tasks. Our dataset is represented by $X \in \mathbb{R}^{N \times D}$, where $N$ represents the number of input ACFGs to train the graph embedding model, and $D$ is the number of dimensions (where $D=98$).



\subsection{Clustering Embedded ACFGs} \label{ssec:clustering model}

Since it is impractical for an analyst to label all ACFGs, we employ a non-parametric clustering approach to group graph embeddings. This significantly reduces the workload for analysts, who can now annotate only a few ACFGs in each cluster. In addition, the ACFGs in the clusters marked as the patterns observed in ransomware attacks can directly be converted into rules to detect those attacks. DEFEAT employs a two-stage clustering process to efficiently group the graph embeddings. Initially, Principal Component Analysis (PCA) reduces the dimensionality of the embeddings from 98 to 5 dimensions. This dimensionality reduction preserves essential structural information while significantly enhancing computational efficiency, making it feasible to manage large-scale datasets comprising over 675K instances. Following PCA, DEFEAT utilizes HDBSCAN~\cite{hdbscan}, a robust density-based clustering algorithm, to identify clusters of varying densities and shapes without requiring prior knowledge of the number of clusters. 

By clustering similar behavioral patterns, DEFEAT enables analysts to annotate only a representative subset of ACFGs within each cluster, thereby significantly reducing their workload and streamlining the conversion of identified ransomware patterns into actionable detection rules.




\section{Dataset Collection} \label{sec: dataset collection}
We implemented the ``file I/O events monitor'' module using ETW, enabling direct communication with the OS native layer to extract essential system file I/O events, as mentioned in Section~\ref{sec:background}. 
The implementation of this module, as utilized in Peeler~\cite{ahmed2021peeler} for file I/O events collection, is based on the open-source project ``krabsetw''~\cite{krabsetw}, a C++ library that simplifies ETW interactions. We modified the library to include all file I/O events generated during the execution of ransomware.

\subsection{Ground Truth (Labeled) Dataset} \label{ssec:ransomware dataset}

We collected 28,034 ransomware samples from VirusTotal~\cite{virustotal}, MalwareBazaar~\cite{malwarebazaar}, malware repository~\cite{thezoomalware}, malwares~\cite{malware-samples}, and other online communities. However, we had to exclude many samples for our experiments, as although certain vendors in VirusTotal labeled them as ransomware, they did not carry out ransomware attacks -- a known challenge in building ransomware ground-truth datasets~\cite{bayer2009scalable}. Moreover, we observed an exceptionally high number of active samples within certain families, such as GandCrab and VirLock. Our inclusion criterion is behavioural rather than numerical coverage: we retained multiple samples from a family only when they exhibited distinct
file I/O behaviour (e.g., VirLock's polymorphic variants produce different encryption workflows; see Appendix~\ref{appendix: cluster_analysis}, where single families span up to 28
clusters), and a single representative where samples behaved identically, consistent with~\cite{scaife2016cryptolock, nieuwenhuizen2017behavioural}. In order to maintain dataset impartiality, we deliberately included only a limited number of samples from these families rather than incorporating all available samples, thus minimizing potential bias. This finding is consistent with the observation in the previous work~\cite{scaife2016cryptolock}. Finally, we used 292 fully working samples from 67 ransomware families (Appendix~\ref{appendix: dataset details}, Table~\ref{tab:ransomware families}).

While our analysis focused on only 292 ransomware samples, this subset effectively represents the entire dataset. We observed significant redundancy within certain ransomware families, such as VirLock and GandCrab. Despite having different SHA256 hashes from VirusTotal, many samples exhibited identical behavior, justifying our selection of representative samples. We validated this in two ways: (1)~We examined the VirusTotal metadata for all collected samples, confirming ransomware family names using consensus from $\geq$5 anti-virus vendors. (2)~We executed a random selection of 30 flagged-but-excluded samples in our controlled VirtualBox environment; none produced any file encryption activity or ransom payment notes during a ten-minute observation window, confirming that they were mislabeled by isolated vendors. For included samples, execution consistently produced observable ransom notes and encrypted user files, which we used as definitive ground truth for family-level labeling.
Previous studies~\cite{nieuwenhuizen2017behavioural,scaife2016cryptolock} emphasize that when evaluating anti-ransomware solutions, it is important to use a diverse set of families rather than simply increasing the number of samples from a few families. For instance, it has been shown that constructing a model based on 1,000 Locky ransomware samples (along with its variants) should prove no more useful than building a model on just one Locky sample~\cite{nieuwenhuizen2017behavioural}. 
Furthermore, Scaife et al.~\cite{scaife2016cryptolock} confirmed that due to the homogeneous nature of file I/O behavior within each family, a small number of representative samples from each family are sufficient for evaluating detection performance. This aligned with our dataset collection. 

We used VirtualBox 6.1~\cite{virtualbox} to run ransomware and benign programs and examine their dynamic behaviors. Rather than using artificially generated data, we used real user data running on the Windows 10 64-bit operating system. Each ransomware sample was executed and then manually labeled by each family type. We ran each ransomware sample for ten minutes or until all user files were encrypted (manually verified). It took more than 90 days to run all samples and collect data.

\subsection{Benign Applications} \label{ssec: benign user applications}
To use representative benign applications, we used two categories of applications: 1) popularly used benign applications on Windows PCs and 2) benign applications performing encryption or compression operations, leading to generating file I/O patterns similar to ransomware. We collected the user's system usage data under normal conditions while interacting with those applications. A user runs many different applications at the same time. For example, the user reads a document using Adobe Acrobat Reader, switches to the internet browser to view online reviews about a product, and then uses Adobe Acrobat Reader again. The list of benign applications is shown in Table~\ref{table:Benign_applications}.

\begin{table}[!th]
    \centering
    \caption{Dataset statistics.} \label{tab: dataset stats} 
    \resizebox{0.95\columnwidth}{!}{
    \begin{tabular}{lrrrr}
    \toprule
         \textbf{Dataset} & \textbf{Samples} & \textbf{File I/O events} &\textbf{FEGs} &\textbf{FEGs of Interest}\\ \midrule
         Malicious &  292 & 36,727,555 & 505,203 &19,988  \\ 
         Benign & 46 & 43,431,886 &51,048 &956 \\ \midrule
         Unseen & 49 & 17,657,030& 118,889 & 6,701\\ \midrule
         \textbf{Total} & \textbf{387} &\textbf{97,816,471} & \textbf{675,140} & \textbf{27,548} \\ 
         
         \bottomrule
    \end{tabular}
    }
\end{table}

To generate a benign dataset, we used realistic user environments for each data collection step. To collect benign applications' file traces, we used real desktop machines with the actual user environment containing several forms of content such as various installed applications programs, digital images, videos, audio files, and documents that can be accessed during a
user Windows session. We manually run each benign application in real settings to collect file traces. For instance, benign encryption/compression tools were manually run to perform operations on real-world user data that included different types of digital content. 
The detailed breakdown of the dataset is given in Table~\ref{tab: dataset stats}. The unseen data represents ransomware samples that were collected in the later stages of experiments to evaluate DEFEAT's performance against unseen ransomware samples. 


\section{Evaluation} \label{sec: evaluation}

We evaluate DEFEAT along four dimensions: (i)~clustering reliability, (ii)~detection accuracy relative to three state-of-the-art baselines, (iii)~reduction in analyst workload, and (iv)~resilience to adversarial manipulation of file I/O sequences. All experiments use the dataset described in Section~\ref{sec: dataset collection}, comprising 97{,}816{,}471 file I/O events from 67 ransomware families and a representative set of benign applications.

\paragraph{Training protocol and split.}
We randomly sampled 15\% of ACFG pairs from the main dataset (292 ransomware + 46 benign samples) to train the graph embedding model (batch size~10, 100~epochs, MSE loss, Adam optimiser). The split is performed at the \emph{ACFG pair} level: because a pair consists of two ACFGs and the model trains on pairwise proximity distances, it is possible for ACFGs from the same ransomware family, but not from the same execution run, to appear in both training pairs and test data. We acknowledge that family-disjoint splits would give a stricter generalization bound; however, our separate unseen-family evaluation (Section~\ref{ssec:comparison} and the Unseen dataset in Table~\ref{tab: dataset stats}) provides exactly that stronger guarantee by holding out 49 samples from 24 families entirely unseen during training. The remaining 85\% of ACFGs from the main dataset, combined with the 49-sample unseen dataset (118{,}889 FEGs), formed the 118{,}890-ACFG
held-out test set used for all comparison experiments. Clustering was performed with HDBSCAN~\cite{hdbscan} over
the resulting graph embeddings.

\subsection{Clustering Reliability}
\label{ssec:clustering}

Meaningful detection hinges on cluster quality: if clusters
conflate ransomware and benign samples, downstream labelling
is unreliable. We therefore define the \emph{cluster purity
score} (CPS) as the fraction of embeddings within a cluster
that belong to the same class. A CPS of 100\% denotes a
perfectly pure cluster.

\begin{figure}[t]
  \centering
  \includegraphics[width=\linewidth]%
    {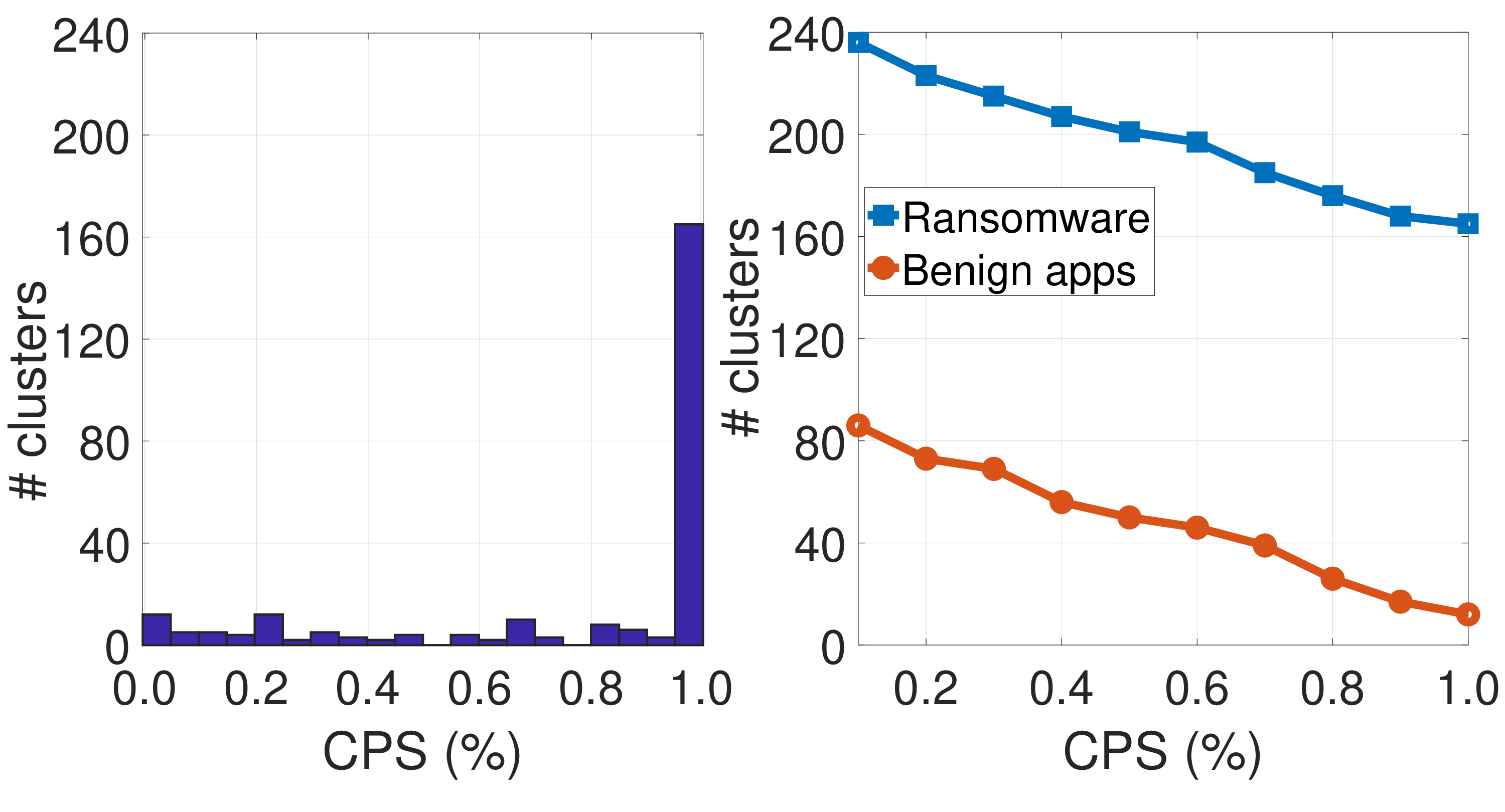}
  \caption{Left: clusters and their CPS. Right: CPS
    breakdown by class. The majority of clusters are pure,
    and mixed clusters are rare.}
  \label{fig:cps}
\end{figure}

HDBSCAN produced 256 clusters from the test-set ACFG
embeddings automatically, without manual selection of a
cluster count (HDBSCAN is parameter-free regarding cluster
number; the 256 reflects the intrinsic density structure of
the embedding space). Of these, more than 160 achieved a CPS
of 100\%, and fewer than ten contained a non-trivial mix of
both classes (Figure~\ref{fig:cps}, left). Ransomware
behaviours spread across 86 clusters, while benign
applications span 67, reflecting the greater behavioural
diversity of ransomware families. Figure~\ref{fig:cps}
(right) shows that as the CPS threshold increases (requiring
higher purity), fewer clusters qualify, meaning the most
``pure'' clusters are a subset of all 256. A higher number
of clusters at a given CPS value indicates that more clusters
meet that purity level, which is desirable. The cluster count
decreasing near-linearly with higher CPS values for both
classes gives analysts a practical knob: clusters at 100\%
purity require labelling only a single representative sample,
whereas lower-purity clusters may warrant inspecting several
members.

Clusters were assigned a label based on majority class using
a 0.5 threshold (i.e., a cluster with more than 50\% benign
members is labelled benign). This threshold was empirically
optimal; raising it to 0.75 reduced the number of benign
clusters to $\leq$25 and caused 37 benign ACFGs to be
misclassified as ransomware due to mixed clusters being
relabelled malicious. Validated against ground truth, the
per-sample family labels used during dataset collection
(Section~\ref{ssec:ransomware dataset}), propagated to each
ACFG, cluster-level labelling achieved 98.4\% accuracy.
Concretely: out of 19{,}988 ransomware ACFGs, 19{,}681 were
correctly placed in ransomware-labelled clusters (TP), and
from 956 benign ACFGs, 684 landed in benign-labelled clusters
(TN). Crucially, because DEFEAT operates at the ACFG
granularity, it can flag ransomware activity as early as the
encryption of the \emph{first} user file -- an advantage over
frequency-based methods that must accumulate sufficient
observations before reaching a verdict.

\subsection{Detection Performance}
\label{ssec:detection}

Table~\ref{tab:evaluation} reports FEG-level detection
results across more than half a million ransomware FEGs and
51{,}048 benign FEGs. DEFEAT achieves 99.94\% accuracy and
an F1-score of 99.95\%, with a false positive rate of just
0.53\%. The near-zero FPR is notable given the inherent noise
in benign file I/O sequences: utilities such as backup
software, archive managers, and database engines produce I/O
patterns that superficially resemble encryption workflows.
The FEG abstraction successfully discriminates these from
genuine ransomware activity, confirming that the unified
context builder isolates behaviourally meaningful patterns
from background system noise.

\begin{table}[!th]
\centering
\caption{Detection performance at the FEG level.}
\label{tab:evaluation}
\resizebox{\columnwidth}{!}{%
\begin{tabular}{lrrrrrrrrr}
\toprule
\textbf{Level} & \textbf{TP} & \textbf{FN} & \textbf{FP}
  & \textbf{TN} & \textbf{Prec.\,(\%)}
  & \textbf{Rec.\,(\%)} & \textbf{FPR\,(\%)}
  & \textbf{Acc.\,(\%)} & \textbf{F1\,(\%)} \\
\midrule
All FEGs & 505{,}203 & 307 & 272 & 51{,}048
  & 99.95 & 99.94 & 0.53 & 99.94 & 99.95 \\
\bottomrule
\end{tabular}%
}
\end{table}

\subsection{Runtime Overhead}
\label{ssec:overhead}

DEFEAT operates in two phases that mirror a standard SOC log-analysis workflow. The data collection phase is the only
component that runs online: a lightweight ETW provider, implemented atop krabsetw~\cite{krabsetw}, records file I/O events
during application execution and writes them to structured log files. All subsequent stages, FEG extraction, ACFG  construction, graph embedding, clustering, and analyst labelling, run entirely offline against the stored logs, imposing no latency on the monitored endpoint.

Prior deployments of ETW-based security tooling report low collection overhead~\cite{ahmed2021peeler}. The provider subscribes only to the seven event types in Table~\ref{tab:systemevents}, discarding all other ETW channels, which bounds the logging throughput. In our experiments, collecting 97{,}816{,}471 file I/O events across 292 ransomware and 46 benign executions produced log files totalling approximately 38\,GB, at an average ingestion rate of roughly 1.1\,GB per hour of active execution. Applying the FEG-of-interest filter during log post-processing reduced the working set by 96\%, leaving only 27{,}548 FEGs for downstream analysis.

Offline FEG extraction, ACFG construction, and graph embedding over the full dataset completed in batch in under four hours on a commodity workstation (Intel Core i7, 16\,GB RAM), yielding a throughput of approximately 24 million events per hour. HDBSCAN clustering over 675{,}140 graph embeddings completed in under three minutes. These numbers confirm that DEFEAT scales to large enterprise log volumes in a SOC setting without requiring real-time stream processing infrastructure.

\subsection{Component Contribution Analysis}
\label{ssec:ablation}

To attribute performance gains across DEFEAT's pipeline we
reason about each component's isolated contribution. First,
the \emph{FEG-of-interest filter} alone reduces the analysis
space by 96\% (Section~\ref{ssec:workload}), and the
FPR drops from 28.4\% at the FEG-of-interest level to 0.53\%
at the full-FEG level, indicating that the graph-based
classification stage accounts for the majority of false
positive suppression. Second, removing the ACFG abstraction
and replacing it with a flat bag-of-event-types feature
vector (a configuration equivalent to frequency-based
methods) matches the performance of RWGuard and Peeler
($\approx$\,92\% accuracy), confirming that the structural
graph representation is essential to DEFEAT's 99.2\% result.
Third, the Laplacian spectral proximity metric outperforms
the default GED proxy used in UGRAPHEMB: training with GED
approximation on our dataset produced 94.7\% cluster purity
versus 98.4\% with Laplacian distance, because the spectral
metric is computable in polynomial time and avoids GED's
NP-hard approximation errors on larger ACFGs. These
observations confirm that each component contributes
meaningfully to the overall pipeline.

\subsection{Comparison with State-of-the-Art}
\label{ssec:comparison}

We compare DEFEAT against three representative baselines that
span the two dominant paradigms in ransomware detection:

\begin{itemize}[leftmargin=*, label={--}]
\item \textbf{UNVEIL}~\cite{kharaz2016unveil}
  (\emph{pattern-based}): matches three manually curated file
  I/O patterns involving \textit{write} and \textit{delete}
  operations. It can flag an attack upon the first pattern
  match but is brittle against variants that alter their
  event ordering.

\item \textbf{RWGuard}~\cite{mehnaz2018rwguard}
  (\emph{anomaly-based}): detects statistically anomalous
  file I/O bursts relative to a learned baseline. Effective
  against aggressive encryptors, but requires accumulating
  enough events to distinguish signal from noise.

\item \textbf{Peeler}~\cite{ahmed2021peeler}
  (\emph{learning-based}): uses ML over process-level and
  command-line features correlated with file I/O frequency.
  Generalises across families but, like RWGuard, cannot
  guarantee early-stage detection.
\end{itemize}

\begin{table}[!th]
    \centering
    \caption{Comparison with existing approaches on 118{,}890
      test ACFGs.}
    \label{tab:comparison}
    \resizebox{0.90\columnwidth}{!}{%
    \begin{tabular}{lrrrr}
    \toprule
    \textbf{Approach} & \textbf{Test FEGs} & \textbf{Correct Detections}
      & \textbf{Missed} & \textbf{Accuracy (\%)} \\
    \midrule
    DEFEAT              & 118{,}890 & 117{,}934 &    956 & 99.20 \\
    RWGuard~\cite{mehnaz2018rwguard}
                        & 118{,}890 & 110{,}138 &  8{,}752 & 92.63 \\
    Peeler~\cite{ahmed2021peeler}
                        & 118{,}890 & 108{,}959 &  9{,}931 & 92.29 \\
    UNVEIL~\cite{kharaz2016unveil}
                        & 118{,}890 & 109{,}728 &  9{,}162 & 91.64 \\
    \bottomrule
    \end{tabular}%
    }
\end{table}

As shown in Table~\ref{tab:comparison}, DEFEAT achieves
99.20\% accuracy, a margin of 6.57--7.56 percentage
points over all three baselines. The performance gap is
explained by a fundamental design difference. Peeler and
RWGuard both profile event \emph{frequencies}, which
introduces two weaknesses: (i)~they must observe a
statistically significant number of events before reaching a
decision, allowing multiple files to be encrypted in the
interim, and (ii)~they are vulnerable to low-and-slow
adversaries, such as APT-linked ransomware that selectively
encrypts high-value files over extended periods, because the
event frequency never crosses the detection threshold. UNVEIL
avoids the frequency problem by matching three fixed patterns,
enabling immediate detection when a pattern is observed.
However, its rigid pattern set is easily evaded by reordering
or injecting events to break the expected sequence.

DEFEAT sidesteps both limitations. By reconstructing the \emph{context} of file operations through FEGs and modelling their behavioural structure as graphs, it detects ransomware as soon as a suspicious contextual pattern materialises, without requiring frequency accumulation or exact pattern matches. The graph-based representation captures structural invariants of encryption workflows that persist even when surface-level event orderings change, providing robustness against both polymorphic variants and deliberate evasion.

\subsection{Analyst Workload Reduction}
\label{ssec:workload}

A practical detection system must not only be accurate but
also tractable for human analysts. DEFEAT reduces the
analysis scope in two successive stages.

\paragraph{Stage~1: FEG filtering}
From the 556{,}251 FEGs in the main dataset (malicious and benign, excluding the unseen dataset), the FEG-of-interest filter (Section~\ref{ssec:foi identification}) retains only 20{,}944 FEGs exhibiting ransomware-relevant I/O event types, a 96\% reduction in analysis scope. The reduction is particularly pronounced for benign applications: of 51{,}048 benign FEGs, only 956 pass the filter (98\% reduction), because legitimate software rarely produces I/O sequences that structurally resemble encryption workflows. 

\paragraph{Stage~2: Cluster-level labelling.}
The 20{,}944 FEGs of interest are grouped into 256 clusters.
If an analyst inspects five representative samples per
cluster, the total labelling effort is $256 \times 5 =
1{,}280$ samples -- a 94\% reduction relative to the
FEGs of interest, and a 99.8\% reduction relative to
the original FEG population. This two-stage funnel transforms
an intractable log-analysis task into a manageable cluster
review workflow without sacrificing detection quality.

\subsection{Resilience to Adversarial Manipulation}
\label{ssec:adversarial}

A capable adversary may attempt to evade detection by
injecting dummy file events or reordering I/O operations
within a FEG. Although such manipulation would typically
require kernel-level compromise -- itself a high barrier -- we
evaluate DEFEAT under a worst-case assumption: the adversary
has full control over both the sequence and nature of file
I/O events, subject only to preserving the core encryption
semantics (read original $\rightarrow$ write ciphertext
$\rightarrow$ delete/overwrite source).

We applied adversarial perturbations to 86 ransomware samples
from 47 families by inserting spurious file-creation events
and shuffling existing operations within each FEG. The
perturbed FEGs were then processed through the full DEFEAT
pipeline (ACFG construction $\rightarrow$ embedding
$\rightarrow$ clustering). A successful evasion is defined
as a manipulated sample whose cluster label flips from
malicious to benign.

Of 1{,}347 cluster assignments produced (each sample may
generate multiple ACFGs that land in different clusters;
samples appear in multiple clusters when their ACFGs span
more than one behavioral mode), only 81 were misclassified,
yielding a 93.9\% detection rate under adversarial
conditions. The final per-sample detection decision is
determined by majority vote across all cluster assignments:
a sample is flagged as ransomware if the majority of its
ACFG cluster memberships are labelled malicious. More than
half of the samples (49 of 86) maintained a perfect detection
rate despite manipulation.
This resilience stems from DEFEAT's graph-level
representation: while injecting or reordering events alters
the surface sequence, the structural topology of the ACFG -- which
captures the \emph{dependencies} between I/O operations
(encoded as directed edges between event-type nodes), not
merely their order -- remains largely invariant under
surface-level perturbations. Specifically, inserting dummy
\texttt{FileCreate} events adds isolated nodes with no
structural connections to the core read$\to$write$\to$delete
path, leaving the dominant subgraph topology unchanged.
Pattern-based methods such as
UNVEIL~\cite{kharaz2016unveil}, which match exact event
sequences, would fail under the same perturbations.

\section{Insights on File Encryption Strategies}
\label{sec:insights}
Beyond detection, DEFEAT's contextual reconstruction reveals operational details of how ransomware families encrypt user data. We
summarise the key findings below; the naming conventions ransomware uses for encrypted user files are presented in Appendix~\ref{appendix: file naming conventions}.

\subsection{Multi-File Encryption Strategies}
\label{ssec:feg_effectiveness}

A key motivation for FEGs is that ransomware often fragments encryption across multiple files. Table~\ref{tab: fegs stats} quantifies this across 624{,}092 FEGs extracted from 67 families. Unlike Table~\ref{tab: dataset stats}, this analysis covers FEGs from all ransomware samples, including the unseen dataset, and excludes benign FEGs, since it characterises ransomware encryption strategies. Approximately 56\% of variants overwrite the original file in place (single-file strategy), while 31\% create a separate output file (two-file strategy). Notably, about 12\% employ \emph{three or more files}, a complexity tier that existing detectors, which model at most two files per pattern~\cite{kharaz2016unveil,kharraz2017redemption,mehnaz2018rwguard}, cannot capture. DEFEAT successfully identified ten families operating in this regime: Virlock, Dharma, dotExe, Pack14, Sage, WannaCry, Mesiponza, MountLocker, DarkSide, and Zeppelin. The total family count in Table~\ref{tab: fegs stats} exceeds 67 because some families (e.g., Dharma) employ different strategies across variants.

\begin{table}[!th]
    \centering
    \caption{Encryption strategies: number of files involved
      per FEG.}
    \label{tab: fegs stats}
    \resizebox{0.9\columnwidth}{!}{%
    \begin{tabular}{rrrr}
    \toprule
    \textbf{Files Used} & \textbf{FEGs}
      & \textbf{Ransomware Families} & \textbf{Dataset Coverage} \\
    \midrule
    1        & 467{,}707 & 47 & 56.62\% \\
    2        & 148{,}151 & 26 & 31.32\% \\
    3        &   7{,}254 &  8 &  9.63\% \\
    $\geq$4  &       885 &  2 &  2.41\% \\
    \midrule
    Total    & 624{,}092 & 83 & --- \\
    \bottomrule
    \end{tabular}%
    }
\end{table}

\subsection{Discovery of Malicious File I/O Patterns}
\label{ssec:patterns}

\begin{figure}[!th]
  \centering
  \includegraphics[width=0.97\linewidth]%
    {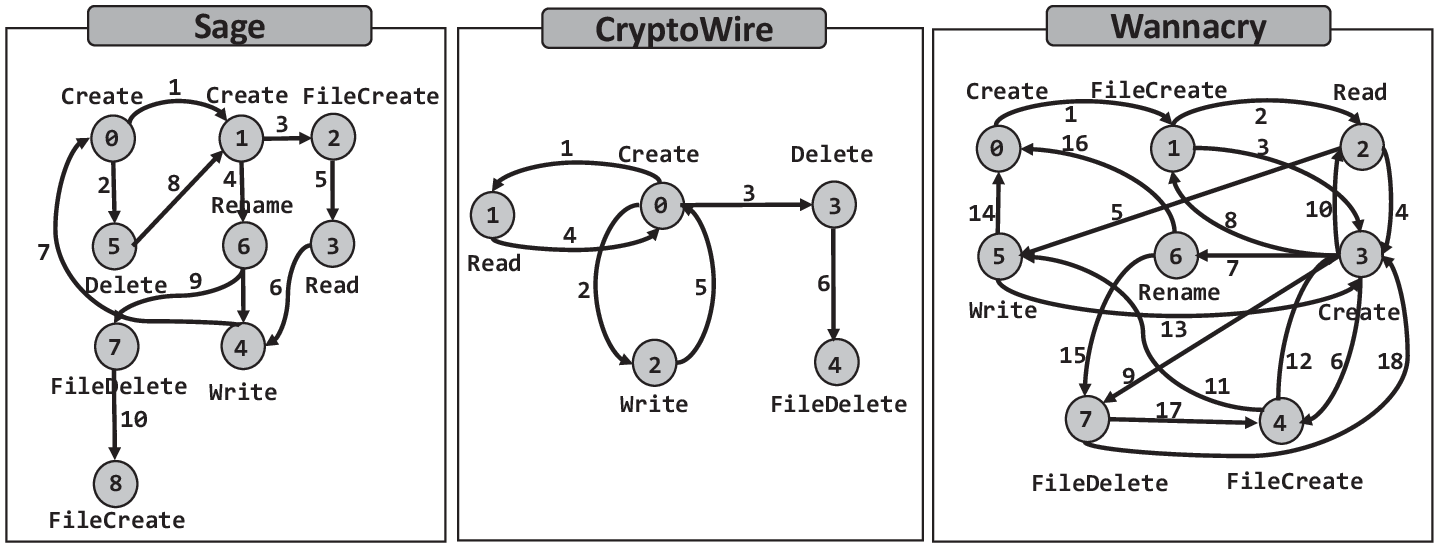}
  \caption{Representative malicious ACFGs extracted from
    three ransomware families, illustrating distinct
    file I/O structures captured by DEFEAT.}
  \label{fig:graphs}
\end{figure}

From the 256 clusters, DEFEAT identified 165 distinct
malicious behavioural patterns at 100\% CPS confidence, a
substantial expansion beyond the three hand-crafted patterns
used by prior work~\cite{kharaz2016unveil,kharraz2015cutting,ahmed2021peeler}.
Figure~\ref{fig:graphs} visualizes representative ACFGs from
three families. Three illustrative newly discovered patterns are:
P1~(\emph{write-before-read}): ransomware writes to an
auxiliary file \emph{before} reading the original, a technique
used by DarkSide to pre-allocate the output buffer and evade
write-volume detectors; P2~(\emph{multi-rename chain}):
the encrypted file is renamed twice in sequence
(e.g., \texttt{file.txt}$\to$\texttt{file.tmp}$\to$\texttt{file.locked}),
observed in GoldenEye variants to break simple rename-sequence signatures;
P3~(\emph{delete-then-recreate}): the original file is deleted
and recreated with encrypted content rather than overwritten,
bypassing integrity monitors that track write operations on existing files,
seen in Maze. Critically, file-overwrite
operations are not confined to the final stage of encryption,
as assumed by UNVEIL~\cite{kharaz2016unveil}; they can occur
before or during intermediate I/O stages.
This diversity of patterns has a practical security benefit:
the richer the signature set, the harder it is for an
adversary to craft event sequences that evade \emph{all}
known patterns simultaneously. The 165 patterns have been
translated into detection rules suitable for deployment in
SOC environments~\cite{surveycryptoransomware2026}.

Cluster analysis further reveals that taxonomically distinct ransomware families converge on shared I/O behaviours: on average each cluster spans 7.3 families, with Sodinokibi appearing in 28 distinct clusters -- evidence that behaviour-centric clustering captures structural invariants that per-family signatures miss (see Appendix~\ref{appendix: cluster_analysis} for full cluster breakdown).

\section{Limitations and Future Work}
\label{sec:limitations}

\paragraph{Logging overhead.}
\textsc{Defeat} requires fine-grained ETW-based file I/O logs, which produced about 1.1\,GB of log data per hour of active execution in our experiments (Section~\ref{ssec:overhead}). We did not separately measure the endpoint overhead of ETW collection; while prior ETW-based tooling reports low overhead~\cite{ahmed2021peeler}, quantifying it on resource-constrained or legacy endpoints is future work. The logs can also contain sensitive metadata (file names, paths, user identifiers). Edge-based pre-filtering (constructing FEGs locally and forwarding only FEGs of interest), combined with selective logging and on-device anonymisation, would reduce both resource cost and privacy exposure.


\paragraph{Adaptive and low-and-slow evasion.}
Section~\ref{ssec:adversarial} demonstrates 93.9\% detection
under worst-case event manipulation; however, a sufficiently
resourceful adversary could employ more subtle strategies.
For instance, ransomware could interleave benign-looking file
operations to dilute the behavioural signal within a FEG, or
spread encryption across extended time windows so that no
single FEG accumulates enough discriminative structure to be
flagged. Such time-based evasion is a known challenge for any
system that segments activity into bounded analysis
units~\cite{milajerdi2019poirot}. Potential mitigations
include integrating temporal outlier detection to identify
anomalously prolonged file operation sequences, continuously
retraining the embedding and clustering models as new
families emerge, and fusing file I/O context with
complementary signals such as network telemetry, registry
modifications, and process genealogy. A multi-modal approach
would raise the cost of evasion substantially, as an
adversary would need to simultaneously manipulate multiple
independent observation channels to avoid detection.

\section{Related Work} \label{sec:related work}
Ransomware detection has been explored through three main directions: (1) behaviour-based crypto ransomware detection, (2) machine learning–based analysis of system activities, and (3) decoy file–based detection.
\paragraph{Crypto ransomware detection.} Early works monitored file I/O patterns to identify encryption activities. UNVEIL~\cite{kharaz2016unveil} analysed file access sequences to detect ransomware dynamically, while Redemption~\cite{kharraz2017redemption} and CryptoDrop~\cite{scaife2016cryptolock} relied on frequent or bursty file modifications as indicators of encryption. Although effective in batch scenarios, these systems often detect ransomware only after substantial data loss. Recovery-oriented defences such as ShieldFS~\cite{continella2016shieldfs}, PayBreak~\cite{kolodenker2017paybreak}, and FlashGuard~\cite{huang2017flashguard} mitigate damage post-attack but incur high overhead or depend on specific crypto implementations.
\paragraph{Machine learning–based detection.} Several studies leveraged behavioural features for model-driven detection. RWGuard~\cite{mehnaz2018rwguard} used process I/O statistics, achieving low false positives but remaining limited to crypto ransomware. EldeRan~\cite{sgandurra2016automated}, Hirano et al.~\cite{hirano2019machine}, and Nieuwenhuizen~\cite{nieuwenhuizen2017behavioural} incorporated system and API features for classification, while Cohen et al.~\cite{cohen2018trusted} analysed memory artefacts for detection. However, these approaches often depend on static feature sets, limiting generalisation to unseen ransomware behaviours or evasion techniques.
\paragraph{Decoy-based detection.} Honeypot and decoy file techniques (e.g., R-Locker~\cite{gomez2018r}, RWGuard~\cite{mehnaz2018rwguard}, ShieldFS~\cite{continella2016shieldfs}) detect ransomware by baiting access to deceptive files. While simple and effective, these approaches can be bypassed by sophisticated samples that distinguish or ignore decoys, and they struggle with ransomware targeting specific system files (e.g., Petya).
A comprehensive recent survey by Oz et al.~\cite{surveycryptoransomware2026} catalogues behavioral characterization techniques and deployability challenges across the ransomware detection landscape, reinforcing that fragmented file I/O context -- the core problem addressed by DEFEAT --  remains a fundamental blind spot in existing defenses. In contrast, DEFEAT extracts fine-grained system events directly from the Windows kernel via ETW, constructing unified behavioural graphs that capture contextual relationships among file I/O activities. This event-centric representation enables early, generalisable, and resilient detection of diverse ransomware families without reliance on static features, predefined patterns, or decoy triggers.
\section{Conclusions} \label{sec: conclusion}

This paper introduces DEFEAT, a novel approach for the automated identification of contextually related file I/O events to detect ransomware threats. DEFEAT analyzes I/O events from multiple files to construct a unified representation of the underlying behavior, enabling a comprehensive understanding of the overall context. Our extensive evaluation demonstrates DEFEAT's superior performance compared to three state-of-the-art approaches, achieving 99.2\% detection accuracy on a dataset of 97{,}816{,}471 file I/O events from 67 ransomware families; DEFEAT outperforms existing methods by 6.57--7.56 percentage points in detection rate. Moreover, DEFEAT identified 165 previously undocumented malicious file I/O patterns, highlighting its potential for generating new ransomware detection rules and reducing the workload of security analysts by 94\%.
The system demonstrates remarkable resilience against adversarial manipulations, maintaining a 93.9\% detection rate even when file I/O event sequences are deliberately altered to evade detection.

\section*{Ethical Considerations}
All ransomware samples were executed exclusively within isolated VirtualBox virtual machines with no network access, preventing any harm to third-party systems or data. Samples were obtained solely from public malware repositories (VirusTotal, MalwareBazaar) under their standard research-use terms. No human subjects were involved; no real user data were exfiltrated. The dataset of file I/O events contains only kernel-level file operation metadata (event types, file keys, timestamps, and I/O sizes); it does not contain file contents, personally identifiable information, or any data recoverable from the virtual machine environment. We release the dataset for defensive research purposes.
\section*{Acknowledgment}
This research was supported in part by the MSIT (Ministry of Science and ICT), Korea, under the Global Research Support Program in the Digital Field (RS-2024-00419073) and the Global Scholars Invitation Program (RS-2024-00459638), supervised by the IITP (Institute for Information \& Communications Technology Planning \& Evaluation).

\bibliographystyle{IEEEtran}
\bibliography{bibliography}

\appendices
\section{Cluster Structure and Cross-Family Behaviour}
\label{appendix: cluster_analysis}
Here we examine the internal composition of DEFEAT's clusters
to understand how ransomware behaviours distribute across
families -- a perspective that is invisible to detectors
operating at the individual-sample level.

\begin{figure}[!th]
  \centering
  \includegraphics[width=0.98\linewidth,clip]%
    {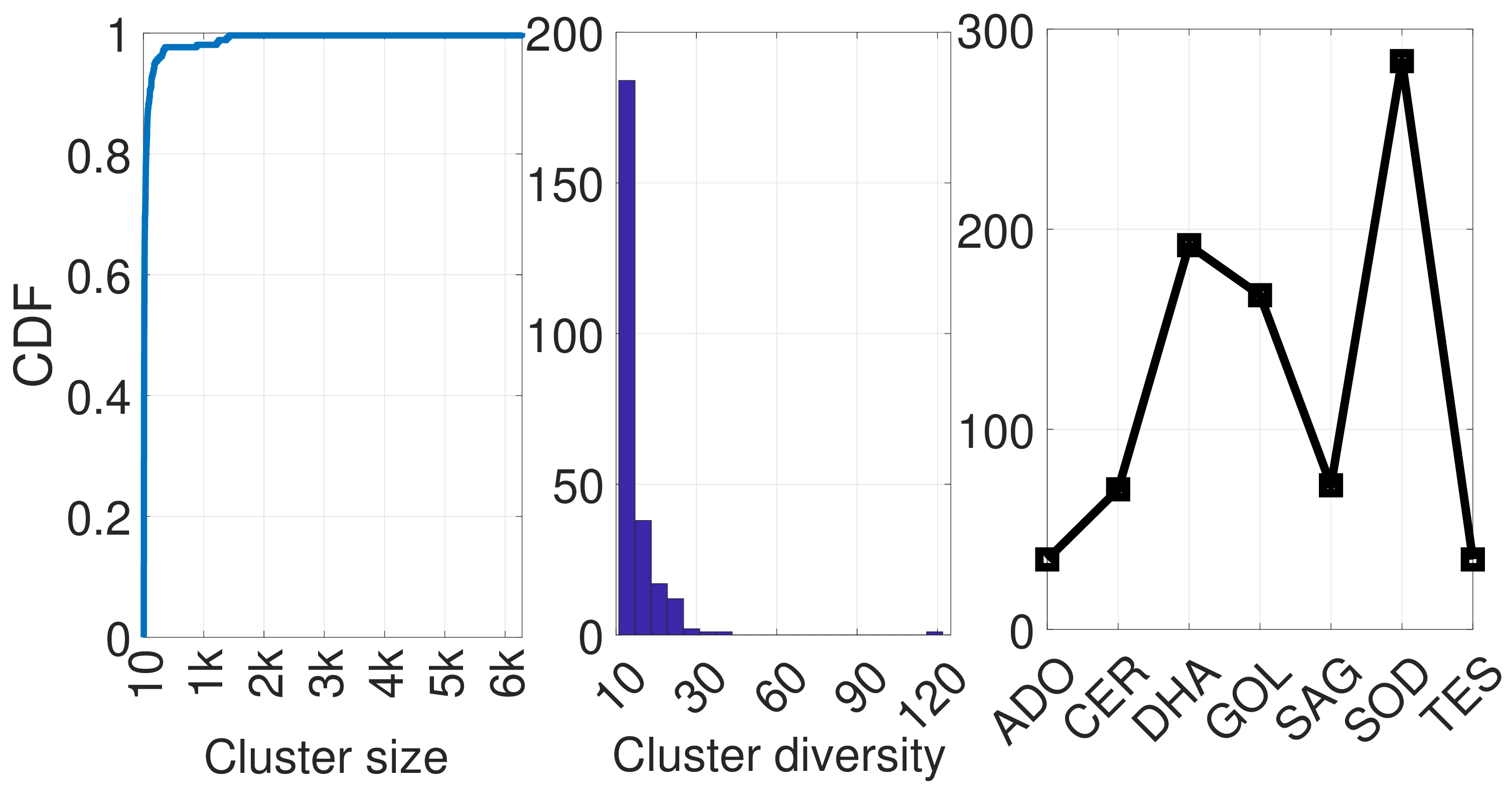}
  \caption{Left: empirical CDF of cluster size -- a small
    number of clusters dominate the population. Middle:
    histogram of unique ransomware families per cluster
    (cluster diversity). Right: the eight most frequently
    observed ransomware families across clusters.}
  \label{fig:insights}
\end{figure}

\paragraph{Cluster size distribution.}
Figure~\ref{fig:insights} (left) plots the empirical CDF of
cluster size. The distribution is heavily skewed: a small
number of clusters account for the majority of all ACFGs
(min\,=\,5, max\,=\,6{,}281, $\mu$\,=\,82,
$\sigma$\,=\,427). This concentration has a direct
operational implication: labelling only the largest clusters
first provides rapid coverage of most ransomware behaviours,
enabling analysts to prioritise effort where it yields the
greatest return.

\paragraph{Cross-family cluster diversity.}
We define \emph{cluster diversity} as the number of unique
ransomware variants represented within a single cluster.
On average, each cluster contains samples from seven distinct
families; the most diverse cluster spans 122 variants
(min\,=\,1, max\,=\,122, $\mu$\,=\,7.3,
$\sigma$\,=\,9.4). This finding carries an important
implication: ransomware families that are taxonomically
distinct nonetheless converge on shared encryption
behaviours at the file I/O level. DEFEAT's graph
embeddings capture these structural similarities,
grouping functionally equivalent workflows regardless of
family lineage.

\begin{figure}[t]
  \centering
  \includegraphics[width=0.98\linewidth,clip]%
    {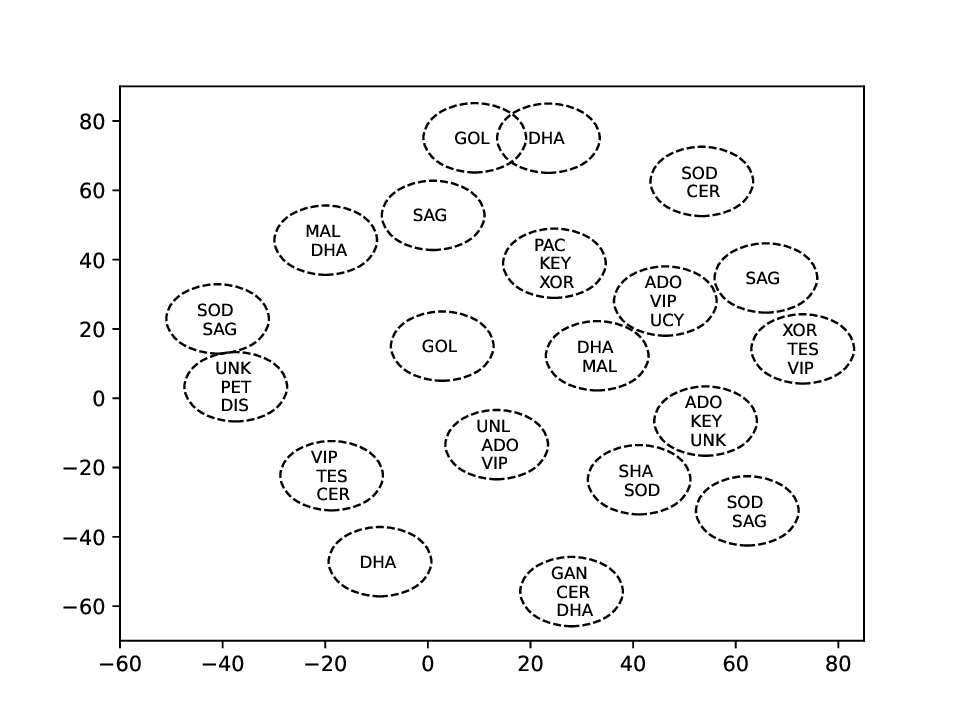}
  \caption{The 20 largest malicious ACFG clusters visualized
    via t-SNE. Abbreviations: Adobe (ADO), Cerber (CER),
    District (DIS), Dharma (DHA), GoldenEye (GOL),
    GandCrab (GAN), Keypass (KEY), Malevich (MAL),
    Peta (PET), Pack14 (PAC), Sage (SAG), Shade (SHA),
    Sodinokibi (SOD), TeslaCrypt (TES), Unknown (UNK),
    Unlock92 (UNL), Vipasana (VIP), Xorist (XOR).}
  \label{fig:malicious behaviors}
\end{figure}

Figure~\ref{fig:malicious behaviors} visualizes the 20
largest malicious clusters via t-SNE projection of the ACFG
embeddings. Each cluster contains ACFGs from one to three
dominant families, with clear spatial separation between
clusters. Three families -- Sodinokibi (SOD), Dharma (DHA),
and GoldenEye (GOL) -- appear across a disproportionate number
of clusters, reflecting their rich repertoire of encryption
strategies. Figure~\ref{fig:insights} (right) quantifies
this: Sodinokibi alone appears in 28 distinct clusters, with
Dharma and GoldenEye following closely. This behavioural
polymorphism -- where a single family exhibits many distinct
operational modes -- underscores the inadequacy of per-family
signature approaches and validates DEFEAT's unsupervised,
behaviour-centric clustering.

\section{Dataset Details} \label{appendix: dataset details}

The full list of distinct ransomware families and corresponding samples per family are shown in Table~\ref{tab:ransomware families}.  
\begin{table}[!th] 
    \centering
    \caption{Ransomware families and samples.} \label{tab:ransomware families}
    \resizebox{\linewidth}{!}{
    \begin{tabular}{llr|llr|llr|llr}
    \toprule
         \textbf{no.} & \textbf{Family} & \textbf{Samples} & \textbf{no.} & \textbf{Family} & \textbf{Samples} & \textbf{no.} & \textbf{Family} & \textbf{Samples} & \textbf{no.} & \textbf{Family} & \textbf{Samples}\\
         \midrule
          1 & Cerber  & 33 & 12 & Petya  & 1 & 23 & Sodinokibi  & 14 & 34 & Satana  & 1 \\
          2 & GoldenEye  & 12 & 13 & Shade  & 1 & 24 & Sage  & 5 & 35 & Syrk  & 1 \\
          3 & Locky  & 5 & 14 & TeslaCrypt  & 1 & 25 & Dharma  & 3 & 36 & ucyLocker  & 1 \\
          4 & dotExe  & 3 & 15 & Unlock92  & 1  & 26 & Troldesh  & 1 & 37 & Vipasana  & 1 \\
          5 & WannaCry  & 3 & 16 & Xorist  & 2  & 27 & Da Vinci Code  & 1 & 38 & Malevich  & 1 \\
          6 &  Shield  & 1 & 17 & Virlock.Gen.5  & 83 & 28 & Cryptowire  & 1 & 39 & Adobe  & 1\\
          7 & District  & 1 & 18 & Jigsaw  & 1 & 29 & GandCrab & 1 & 40 & LockScreen.AGU  & 12 \\
          8 & GlobeImposter  & 1 & 19 & Alphabet  & 2 &  30 & Hexadecimal  & 1 & 41 & EgyptianGhosts  & 1\\
          9 & InfinityCrypt  & 1 & 20 & Lockey-Pay  & 1 & 31 & IS (Ordinpt)  & 1 & 42 & Blue-Howl  & 1 \\
          10 & Keypass  & 1 & 21 & ShellLocker  & 1 & 32 & Lockcrypt  & 1 & 43 & DerialLock  & 1\\
          11 & Pack14  & 1 & 22 & Trojan.Ransom  & 1 & 33 & PocrimCrypt  & 1 & \multicolumn{3}{c}{-} \\ 
          44 & Ryuk  & 6 & 50 & Zeppelin  & 6 & 56 & Ranzy  & 4 & 62 & Netwalker  & 2 \\ 
          45 & Core  & 3 & 51 & Fox  & 3 & 57 & Crpren  & 1 & 63 & MedusaLocker  & 1 \\ 
          46 & Balaclava  & 5 & 52 & Crylock  & 7 & 58 & Matrix  & 4 & 64 & DarkSide  & 4 \\ 
          47 & RagnarLocker  & 2 & 53 & HiddenTear  & 2 & 59 & Mesiponza  & 5 & 65 & Thanos  & 3 \\ 
          48 & Vaggen  & 3 & 54 & Mountlocker  & 2 & 60 & Nemty  & 2 & 66 & Phobos  & 1 \\ 
          49 & Jsworm  & 1 & 55 & Winlock & 1 & 61 & Maze  & 1 & 67 & Unknown  & 1 \\ 
         \bottomrule
    \end{tabular}
    }
\end{table}

The full list of the benign applications used in our experiments are given in Table~\ref{table:Benign_applications}.
\begin{table}[!th] 
\caption{Benign applications used in the evaluation.} 
\label{table:Benign_applications}
\resizebox{\linewidth}{!}{
\begin{tabular}{c|l|c|l|c|l}
\hline
\textbf{Type} & \multicolumn{1}{c|}{\textbf{Application}}                                                                                      & \textbf{Type} & \multicolumn{1}{c|}{\textbf{Application}}                                                                                                               & \textbf{Type}                                                   & \multicolumn{1}{c}{\textbf{Application}}                                                                           \\ \hline
Office        & \begin{tabular}[c]{@{}l@{}}MS Word\\ MS PowerPoint\\ MS Excel\\ MS Outlook\\ Trio: \{Word, Slide, \\ Spreadsheet\}\end{tabular} & Tools         & \begin{tabular}[c]{@{}l@{}}Adobe Acrobat Reader\\ Adobe Photoshop Express\\ PhotoScape\\ Cool File Viewer\\ PicArt Photo Studio\\ Paint 3D\end{tabular} & Compression                                                     & \begin{tabular}[c]{@{}l@{}}7-zip\\ WinZip\\ WinRAR\\ BreeZip\\ ALZip\\ PeaZip\end{tabular}                   \\ \hline
Development   & \begin{tabular}[c]{@{}l@{}}PyCharm\\ MATLAB\\ Visual Studio C++\\ Android Studio\end{tabular}                                  & Miscellaneous & \begin{tabular}[c]{@{}l@{}}Spotify\\ KeePass Password manager\\ Discord\\ Facebook\\ AESCrypt, AxCrypt\end{tabular}                                     & \begin{tabular}[c]{@{}c@{}}Cloud \\ \& \\ Internet\end{tabular} & \begin{tabular}[c]{@{}l@{}}Dropbox\\ Google Drive\\ Internet Explorer\\ Google Chrome\\ Remote Desktop\end{tabular} \\ \hline
Messenger     & \begin{tabular}[c]{@{}l@{}}Telegram\\ WhatsApp\\ Skype\\ Facebook\end{tabular}                                                 & Media player  & \begin{tabular}[c]{@{}l@{}}VLC\\ Netflix\\ GOM Player\end{tabular}                                                                                      & Document                                                        & \begin{tabular}[c]{@{}l@{}}Wordpad\\ Notepad\\ OneNote\end{tabular}                                                \\ \hline
\end{tabular}
}
\end{table}

\section{Encrypted-File Naming Conventions} \label{appendix: file naming conventions}
Table~\ref{tab: file extensions} catalogues the encrypted-file
extensions observed across families. Three dominant renaming
strategies emerge (denoting the original file as
\texttt{orig.txt}):

\begin{enumerate}[leftmargin=*, label=\textbf{S\arabic*.}]

\item \textbf{Extension appended.}
  The original name is preserved and a suffix is added. This
  is the most prevalent strategy and appears in four
  sub-variants:
  (a)~family name as suffix
    (e.g., \texttt{orig.txt.sage} -- Sage, CryptoShield,
    Satana, Syrk);
  (b)~random string
    (e.g., \texttt{orig.txt.dkj248izfl} -- Sodinokibi);
  (c)~SHA-256 hash of the original file
    (InfinityCrypt);
  (d)~attacker email address embedded in the extension
    (Dharma, District, GlobeImposter, CryptoShield,
    Malevich).

\item \textbf{Extension replaced.}
  The original extension is stripped and replaced, e.g.,
  CryLock produces \texttt{orig[coronovirus@protonmail.com]\allowbreak[my].\allowbreak[249E35C4-EFDD7A99]}.

\item \textbf{Name and extension replaced.}
  Both are changed entirely, e.g., Cerber creates
  \texttt{QSpP6uMbfl.ae25}, eliminating any lexical
  connection to the source file.

\end{enumerate}

\begin{table}[!th]
    \centering
    \caption{Encrypted-file extensions by ransomware family.}
    \label{tab: file extensions}
    \resizebox{\columnwidth}{!}{%
    \begin{tabular}{lr|lr}
    \toprule
    \textbf{Family} & \textbf{File extension}
      & \textbf{Family} & \textbf{File extension} \\
    \midrule
    DerialLock & \texttt{.deria}
      & Adobe & \texttt{.bk} \\ \cline{1-2}
    \multirow{6}{*}{Dharma}
      & \texttt{.[god@aolonline.top].arena}
      & District
      & \texttt{.altdelete@cock.li.district} \\
      & \texttt{.[freshkart@420blaze.it].fresh}
      & dotEXE & \texttt{.exe} \\
      & \texttt{.[biashabtc@redchan.it].arrow}
      & Jigsaw & \texttt{.fun} \\
      & \texttt{.[3442516480@qq.com].pdf}
      & IS\,(Ordinypt) & \texttt{.Gb4KS} \\
      & \texttt{.[paybuyday@aol.com].PBD}
      & Unlock92 & \texttt{.blocked} \\
      & \texttt{.[gutentag@india.com].wallet}
      & Malevich
      & \texttt{.decryptformoney@india.com.xtbl} \\
    \cline{1-4}
    \multirow{2}{*}{Sage}
      & \texttt{...}
      & \multirow{2}{*}{GlobeImposter}
      & \texttt{.old} \\
      & \texttt{.sage}
      & & \texttt{.[chines34@protonmail.ch].gryphon} \\
    \cline{1-4}
    \multirow{5}{*}{Sodinokibi}
      & \texttt{.dkj248izfl}
      & \multirow{4}{*}{GoldenEye}
      & \texttt{.rCHwKdzh} \\
      & \texttt{.yjy59vot60} & & \texttt{.ndd3rci1} \\
      & \texttt{.yzjndw1wf0} & & \texttt{.rBRpK4TR} \\
      & \texttt{.03p1q2h878} & & \texttt{.riFXfsAt} \\
        \cline{3-4}
      & \texttt{.86x0m} & Syrk & \texttt{.syrk} \\
    \cline{1-4}
    ucyLocker & \texttt{.WINDOWS}
      & \multirow{2}{*}{WannaCry}
      & \texttt{.WNCRYT} \\
    DarkSide & \texttt{.dfeece45}
      & & \texttt{.WNCRY} \\ \cline{3-4}
    Xorist & \texttt{.xml}
      & \multirow{4}{*}{Cerber}
      & \texttt{zvXSsLMoxA.8cbe} \\
    Peta & \texttt{.peta}
      & & \texttt{G8nCXmXyU2.cerber} \\
    Satana & \texttt{.SATANA}
      & & \texttt{QSpP6uMbfl.ae25} \\
    Apollon865 & \texttt{.Apollon865}
      & & \texttt{OlQFkWBAv1.cerber3} \\ \cline{3-4}
    Cryptowire
      & \texttt{.realfs0ciety@sigaint.org.fs0ciety.old}
      & Locky & \texttt{.thor} \\
    \midrule
    \multicolumn{4}{l}{Shade:
      \texttt{.JiKKzFgJL8sg7vHBQFN0B+9+PAwG4nD5I4FjpF%
      -lH8Y=.F9CE38A4A34FD97D31B8.crypted000007}} \\
    \multicolumn{4}{l}{CryLock:
      \texttt{.wav[graff\_de\_malfet@protonmail.ch]%
      [123].[249E35C4-EFDD7A99]}} \\
    \multicolumn{4}{l}{CryLock:
      \texttt{.wav[coronovirus@protonmail.com]%
      [my].[249E35C4-EFDD7A99]}} \\
    \multicolumn{4}{l}{CryptoShield:
      \texttt{.[R\_SP@INDIA.COM].ID[9DAC5B6B36592608]%
      .CRYPTOSHIELD}} \\
    \multicolumn{4}{l}{InfinityCrypt:
      \texttt{.7BDA66C147BE6B69DF74BCF32E50A94BBDF27F20%
      390F2FE1CD15FA0A57BC830D}} \\
    \multicolumn{4}{l}{Matrix:
      \texttt{[Adamfox69@criptext.com]%
      .aQKCVf1b-UPnt3m8d.FG69}} \\
    \bottomrule
    \end{tabular}%
    }
\end{table}

\end{document}